\documentclass[twocolumn,showpacs,aip,superscriptaddress,reprint,floatfix]{revtex4-1}
\usepackage{graphicx}
\usepackage{dcolumn}
\usepackage{bm}
\usepackage{nicefrac}
\usepackage{hyperref}
\usepackage{gensymb}  
\usepackage{amsmath}
\usepackage{xcolor}
\usepackage{etoolbox}
\usepackage{mathptmx}
\usepackage{amssymb}
\usepackage{soul}
\usepackage{float}

\def\kk         {{\bf k}}
\def\qq		{{\bf q}}
\def\rr         {{\bf r}}

\newcommand{\cinam}{Aix-Marseille Univ., CNRS, CINaM, Centre Interdisciplinaire de Nanoscience de Marseille, UMR 7325, Campus de Luminy, 13288 Marseille cedex 9, France}
\newcommand{\piim}{Aix Marseille Univ, CNRS, PIIM, Physique des Interactions Ioniques et Moléculaires, UMR 7345, Marseille, France}
\newcommand{\cnrmodena}{Centro S3, CNR-Istituto Nanoscienze, I-41125 Modena, Italy}

\newcommand{\RED}[1]{\textcolor{black}{#1}} 
\newcommand{\CLA}[1]{\textcolor{black}{#1}}

\graphicspath{{./}}

\usepackage[utf8]{inputenc}
\usepackage[T1]{fontenc}

\makeatletter
\def\@email#1#2{%
 \endgroup
 \patchcmd{\titleblock@produce}
  {\frontmatter@RRAPformat}
  {\frontmatter@RRAPformat{\produce@RRAP{*#1\href{mailto:#2}{#2}}}\frontmatter@RRAPformat}
  {}{}
}%
\makeatother

\begin{document}
	
\newcommand{\tf}{\textbf}
\newcommand{\bo}{\mathbf}
\newcommand{\br}{{\bf r}}
\newcommand{\be}{\begin{equation}}
\newcommand{\ee}{\end{equation}}
\newcommand{\bea}{\begin{eqnarray}}
\newcommand{\eea}{\end{eqnarray}}
\newcommand{\nup}{n_{\uparrow}}
\newcommand{\ndown}{n_{\downarrow}}
\newcommand{\Id}[1] {\int \! \! {\rm d}^3 #1}
\renewcommand{\v}[1]{{\bf #1}}

\def\qw		{{\bf q},\omega}
\def\ql  	{{\bf q},\lambda}
\def\rw		{{\bf r},\omega}
\def\rt		{{\bf r},t}
\def\rtp	{{\bf r},t'}
\def\tone   {t_{1}}
\def\ttwo   {t_{2}}
\def\taoone   {\tau_{1}}
\def\taotwo   {\tau_{2}}
\def\ttmto   {\tau_{2}-\tau_{1}}
\def\tomtt   {\tau_{1}-\tau_{2}}
\def\pp		{{\bf p}}
\def\rr		{{\bf r}}
\def\rrp	{{\bf r}'}
\def\rrone	{{\bf r}_1}
\def\rrtwo	{{\bf r}_2}
\def\vv		{{\bf v}}
\def\BB		{{\bf B}}
\def\QQ		{{\bf Q}}
\def\GG		{{\bf G}}
\def\RRI        {{\bf R}_{Is}}
\def\RRJ        {{\bf R}_{Js'}}
\def\RR		{{\bf R}}
\def\KK		{{\bf K}}
\def\EE		{{\bf E}}
\def\qq		{{\bf q}}
\def\kk		{{\bf k}}
\def\jj		{{\bf j}}
\def\calGzero   {{\cal G}^{(0)}(\kk,\gon)}
\def\calG       {{\cal G}(\kk,\gon)}
\def\intr	{\int\,d{\bf r}\,}
\def\intrp	{\int\,d{\bf r}'\,}
\def\introne	{\int\,d{\bf r}_{1}\,}
\def\intrtwo	{\int\,d{\bf r}_{2}\,}
\def\intao     {\int\,d\tau\,}

\def\ga         {\alpha}
\def\gb         {\beta}
\def\gc         {\gamma}
\def\gC         {\Gamma}
\def\gd         {\delta}
\def\gD         {\Delta}
\def\gee        {\epsilon}
\def\gl         {\lambda}
\def\gL         {\Lambda}
\def\go         {\omega}
\def\gO         {\Omega}
\def\goql       {\omega_{\qq \gl}}
\def\goqlt       {\omega_{\tilde \qq \gl}}
\def\gql        {\qq \gl}
\def\gon        {\omega_{n}}
\def\goj        {\omega_{j}}
\def\goi        {\omega_{i}}
\def\igon       {i\omega_{n}}
\def\igoi       {i\omega_{i}}
\def\igoj       {i\omega_{j}}
\def\gr         {\rho}
\def\gs         {\sigma}
\def\gS         {\Sigma}
\def\gSnko      {\Sigma_{\nk}(\go)}
\def\Skon       {\Sigma(\kk,\gon)}
\def\Snko       {\Sigma_{n\kk}(\go)}
\def\Snkoi      {\Sigma_{n\kk}(\goi)}
\def\gt         {\theta}
\def\gu         {\tau}
\def\gup         {\tau'}
\def\ek         {\gee_{\bf k}}
\def\ekpq       {\gee_{\bf {k+q}}}
\def\capo       {\right.\\ \left.}
\def\dn         {\downarrow}
\def\up         {\uparrow}
\def\rar        {\rightarrow}  
\def\la         {\langle}
\def\ra         {\rangle}
\def\dg         {\dagger}
\def\grG        {\gr_{nn'}(\kk,\qq,{\bf G})}
\def\grGo       {\gr_{nn'}(\kk,\qq,{\bf G_1})}
\def\grGts      {\gr_{nn'}^*(\kk,\qq,{\bf G_2})}
\def\uob        {\frac {1}{\gb}}
\def\uoZ        {\frac {1}{Z}}
\def\uot        {\frac {1}{2}}
\def\uop        {\frac {1}{\pi}}
\def\uon        {\frac {1}{N_{q}}}
\def\sumon      {\sum_{n=-\infty}^{+\infty}}
\def\sumoj      {\sum_{j=-\infty}^{+\infty}}
\renewcommand{\[}{\left[}
\renewcommand{\]}{\right]}
\renewcommand{\(}{\left(}
\renewcommand{\)}{\right)}
\def\MBPT       {many-body perturbation theory}
\def\MB         {many-body }
\def\GS         {ground state }
\def\GF         {Green's function }
\def\GFv        {Green's function, }
\def\GFs         {Green's functions }
\def\zpme         {zero point motion effect}

\def\efield	{ {\bf {\cal E}}}

\def\FF		    {{\bf F}}

\def\dk         {\frac{d\,{\bf k}}{\(2 \pi\)^3}}
\def\dku        {\frac{d\,{\bf k}_1}{\(2 \pi\)^3}}
\def\dq         {\frac{d\,{\bf q}}{\(2 \pi\)^3}}

\def\ecut       {\it E_{cut} }
\def\tpa        {{\it trans}--polyacetylene }
\def\Tpa        {{\it Trans}--polyacetylene }
\def\tpe        {polyethylene }

\def\Hscf       {H_{scf} }
\def\Vscf       {\widehat{V}_{scf} }
\def\Hzero      {\widehat{H}_0}
\def\Hone       {\widehat{H}_1}
\def\Htwo       {\widehat{H}_2}

\def\epc        {electron--phonon coupling }
\def\epcp        {electron--phonon coupling. }
\def\epcv        {electron--phonon coupling, }
\def\epi        {electron--phonon interaction }
\def\epip        {electron--phonon interaction.}
\def\epiv        {electron--phonon interaction,}
\def\ep         {electron--phonon }
\def\dw         {Debye--Waller }
\def\se         {self--energy }
\def\Se         {Self--energy }
\def\sep         {self--energy. }
\def\sev         {self--energy, }

\def\nk         {n{\bf k}}
\def\npkp       {n' \kk'}
\def\npk        {n' \kk}
\def\rnpk       {\mid  n'    \kk    \ra}
\def\lnpk       {\la   n'    \kk    \mid}
\def\rnk        {\mid  n     \kk    \ra}
\def\lnk        {\la   n     \kk    \mid}
\def\rnpkp      {\mid  n'    \kk'   \ra}
\def\lnpkp      {\la   n'    \kk'   \mid}
\def\rnpkpq     {\mid  n',{\bf k+q} \ra}
\def\npkmq      {n' {\bf k-q}}
\def\npkpq      {n' {\bf k+q}}

\def\uu		{{\bf u}_{Is}}
\def\uais	{u_{\ga,Is}}
\def\ubis	{u_{\gb,Is}}
\def\ubjsp	{u_{\gb,Js'}}

\def\tlim        {T going to zero limit}
\def\bql         {\hat{b}_{\qq \gl}}
\def\bmql        {b_{-\qq \gl}}
\def\bdql        {\hat{b}^{\dg}_{-\qq \gl}}
\def\bdpql       {\hat{b}^{\dg}_{\qq \gl}}
\def\gsq         {{\mid g^{\gql}_{n n' \kk} \mid}^2}
\def\Sf          {spectral function}
\def\Sfs         {spectral functions}
\def\qp          {quasiparticle}

\def\prl         {Phys. Rev. Lett. }
\def\prb         {Phys. Rev. B }
\def\pr          {Phys. Rev. }
\def\rpm         {Rev. Mod. Phys. }
\def\jpc         {J. Phys. Chem. }
\def\jp          {J. Phys. C }
\def\jcp         {J. Chem. Phys. }
\def\pss         {Phys. Stat. Sol. }

\def\self       {\epsfig{figure=Appendix/logo.ps,height=.5cm}\,}
\def\bverb      {\renewcommand{\baselinestretch}{1}\begin{verbatim}}
\def\everb  	{\end{verbatim}\renewcommand{\baselinestretch}{1.3}}

\def\ai         {{\it ab-initio} }
\def\ga         {\alpha}
\def\gb         {\beta}
\def\gC         {\Gamma}
\def\gD         {\Delta}
\def\gee        {\varepsilon}
\def\gl         {\lambda}
\def\gL         {\Lambda}
\def\go         {\omega}
\def\goql       {\omega_{\qq \gl}}
\def\gql        {\qq \gl}
\def\gS         {\Sigma}
\def\la         {\langle}
\def\ra         {\rangle}
\def\kk         {{\bf k}}
\def\qq         {{\bf q}}
\def\Gone       {{\bf G_{1}}}
\def\Gtwo       {{\bf G_{2}}}

\def\epc        {EP coupling }
\def\epcp       {EP coupling. }
\def\epi        {EP interaction }
\def\epip       {EP interaction. }
\def\ep         {EP }
\def\dw         {Debye--Waller }

\def\MB         {many body }
\def\se         {self-energy }
\def\sev        {self-energy, }
\def\sep        {self-energy. }
\def\Sf         {SF}
\def\Sfs        {SFs}
\def\qp         {quasiparticle}
\def\nk         {n\kk}
\def\npkp       {n' \kk'}
\def\npk        {n' \kk}
\def\rnk        {\mid  n     \kk    \ra}
\def\lnk        {\la   n     \kk    \mid}
\def\rnpkp      {\mid  n'    \kk'   \ra}
\def\lnpkp      {\la   n'    \kk'   \mid}
\def\uot        {\frac {1}{2}}

\def\Znk        {Z_{\nk}}
\def\zpme       {zero point motion effect}

\def\zbGN       {{\it zb}--GaN }
\def\ZbGN       {{\it Zb}--GaN }
\def\zbGNv       {{\it zb}--GaN, }
\def\zbGNp       {{\it zb}--GaN. }

\renewcommand{\[}{\left[}
\renewcommand{\]}{\right]}
\renewcommand{\(}{\left(}
\renewcommand{\)}{\right)}
\newcommand{\ltsim}{\protect\raisebox{-0.5ex}{$\:\stackrel{\textstyle <}{\sim}\:$}}

\newcommand{\lv}{{\bf a}}
\newcommand{\hh}{{\bf h}}
\newcommand{\PP}{{\bf P}}
\newcommand{\HH}{{\bf H}}
\newcommand{\SiS}{{\bf \Sigma}}
\newcommand{\VV}{{\bf V}}
\newcommand{\UU}{{\bf U}}
\newcommand{\w}{\omega}
\newcommand{\ex}{\mathrm{e}}
\newcommand{\ben}{\begin{equation*}}
\newcommand{\een}{\end{equation*}}
\newcommand{\bean}{\begin{eqnarray*}}
\newcommand{\eean}{\end{eqnarray*}}
\renewcommand{\v}[1]{{\bf #1}}
\renewcommand{\[}{\left[}
\renewcommand{\]}{\right]}
\renewcommand{\(}{\left(}
\renewcommand{\)}{\right)}
\newcommand{\chitwo}{\chi^{(2)}}
\def\efield{\boldsymbol{\cal E}} 
\def\ket#1{\vert#1\rangle}
\def\bra#1{\langle#1\vert}
\def\susc#1{\chi^{(#1)}}
\def\ket#1{\vert#1\rangle}
\def\bra#1{\langle#1\vert}

\def\ai{\textit{ab initio}}
\def\hbn{\textit{h}-BN}
\def\ks{Kohn-Sham}

	\title{Phonon assisted light absorption and emission in cubic Boron Nitride}
	
	\author{Ashwin Pillai}
	\affiliation{\cinam}
	\author{Elena Cannuccia}
	\affiliation{\piim}
	\author{Aur\'elien Manchon}%
	\affiliation{\cinam}
	\author{Fulvio Paleari}%
	\affiliation{\cnrmodena}
	\author{Claudio Attaccalite}%
	\email{claudio.attaccalite@univ-amu.fr}
	\affiliation{\cinam}
	\date{\today}
	
	\begin{abstract}
            Cubic boron nitride (cBN) is a wide--bandgap polymorph of boron nitride whose optical response remains only partially understood due to the coexistence of indirect electronic transitions and strong exciton--phonon coupling. Using first-principles Many-Body Perturbation Theory (MBPT), we investigate the optical properties of cBN by combining GW quasiparticle corrections with Bethe-Salpeter equation calculations of excitonic effects. Phonon-assisted absorption and emission processes are explicitly included through the exciton--phonon coupling formalism. We find that phonon--mediated optical transitions provide a dominant contribution to both absorption and luminescence spectra, partially reconciling the discrepancy between the theoretical optical gap ($\approx 11$~eV) and experimental emission around 6--7 eV. Our results demonstrate the importance of including exciton--phonon interactions for the correct interpretation of experimental spectra,  offering new insights into light emission in wide-bandgap materials.

	\end{abstract}
	
	\maketitle
	
	
    Boron nitride (BN) exhibits a rich polymorphism, encompassing phases ranging from the common hexagonal (hBN) and cubic (cBN) structures to more complex polytypes.\cite{doi:10.1126/sciadv.aau5832} A significant challenge in studying these materials is the frequent coexistence of different phases and the difficulty in obtaining perfectly pure samples. 
    This complexity underscores the need for precise characterization techniques. While Raman spectroscopy is a widely used characterization tool in materials science, its utility for BN polymorphs is often limited due to the material's few active Raman modes and its deep UV band gap, which force nonresonant conditions. Consequently, luminescence spectroscopy emerges as a far more powerful and informative probe. The sensitivity of luminescence to electronic structure, defects, and stacking order makes it exceptionally well-suited for characterizing BN nanostructures, where it can reveal details that remain inaccessible to Raman scattering\cite{jaffrennou2007origin}. For example, luminescence has proven to be a versatile tool, successfully distinguishing between different stacking sequences in the hexagonal phase\cite{PhysRevLett.131.206902} and probing the effects of strain.\cite{10.21468/SciPostPhys.12.5.145, schue2017proprietes}\\
    The past few years have witnessed intense experimental and theoretical interest in the luminescence properties of BN polymorphs.\cite{cassabois2021hexagonal, Schue2019, elias2021flat} A unifying feature of most BN phases is their indirect band gap, meaning that luminescence typically arises from phonon-assisted transitions. Remarkably, despite this indirect nature, the luminescence efficiency in hBN has been shown to rival that of direct band gap materials.\cite{Cassabois2016, Schue2019} This high efficiency should be a consequence of the strong exciton--phonon coupling inherent to BN systems.\cite{PhysRevLett.122.067401, PhysRevLett.127.137401} Beyond hBN, luminescence has also been investigated in the wurtzite phase (wBN), though a direct comparison between experiment and theory remains challenging due to difficulties in synthesizing pure wBN.\cite{PhysRevMaterials.7.055201} Even in the direct band gap monolayer, luminescence studies are central in the  debates over the origin of emission features, whether from direct or indirect transitions.\cite{rousseau2024spatially, liu2025direct, PhysRevMaterials.7.024006, marini2024optical,Fu2025} \\ 
	Within this family of polymorphs, cubic boron nitride (cBN) presents a particularly compelling and unresolved puzzle.\cite{samantaray2005review,mirkarimi1997review} As a thermodynamically stable phase at standard conditions, cBN has been synthesized primarily by chemical vapor deposition.\cite{mirkarimi1997review, samantaray2005review} Experimental optical measurements, including those under pressure,\cite{onodera1993pressure} consistently report an optical absorption onset around 6--7~eV.\cite{miyata1989optical, galambosi2001nonresonant, sobolev1999optical} However, state-of-the-art theoretical calculations, which typically neglect the electron--phonon interaction, predict a much larger direct optical gap of approximately 11~eV.\cite{PhysRevB.43.9126, cappellini2001optical, satta2004many} This significant discrepancy has been tentatively attributed to phonon-assisted indirect transitions, which could lower the observed absorption edge. Furthermore, the potential presence of hBN inclusions in experimental samples, which luminesce around 6~eV, may mask the intrinsic phonon-assisted emission signal from pure cBN.
	
	From a theoretical point of view, exciton--phonon coupling has been mainly studied with different approaches: by finite difference methods\cite{Cannuccia2019, Paleari2019, PhysRevMaterials.7.055201, 10.21468/SciPostPhys.12.5.145} and the direct calculation of microscopic exciton--phonon scattering amplitudes.\cite{PhysRevMaterials.7.024006, marini2024optical, PhysRevLett.131.206902} In this work, we follow this second strategy to investigate the role of atomic vibrations in the optical response of cBN.

	This work aims to address the long-standing discrepancy between theoretical predictions and experimental optical measurements in cubic boron nitride. While previous studies have investigated the optical properties of cBN either neglecting phonon effects \cite{satta2004many} or treating phonon-assisted absorption without excitonic interactions \cite{iqbal2024phonon}, a unified first-principles description combining excitons and phonon scattering has so far been missing. 
	
	Here, we present a first-principles investigation of the optical absorption and luminescence of cBN that explicitly includes exciton--phonon coupling within a Many--Body Perturbation Theory (MBPT) framework. 
	We assess the role of indirect excitonic transitions in shaping the optical spectra of cBN and clarify their relevance for the interpretation of experimental observations.
	
    This information is critical, given that cBN is a large-gap material where strongly bound excitons are expected and the coupling between excitons and phonons is known to be \RED{crucial} in other BN polymorphs.\cite{PhysRevLett.122.067401, PhysRevLett.127.137401} Furthermore, understanding exciton--phonon coupling in cBN has broader implications, particularly in light of recent studies on small polaron formation in doped cBN systems\cite{shirodkar2023small} and the reported observation of the electron--phonon Fano effect in cBN.\cite{zhu2023electron}\\ 

    	\begin{figure}[ht]
		\centering
		\includegraphics[width=0.50\textwidth]{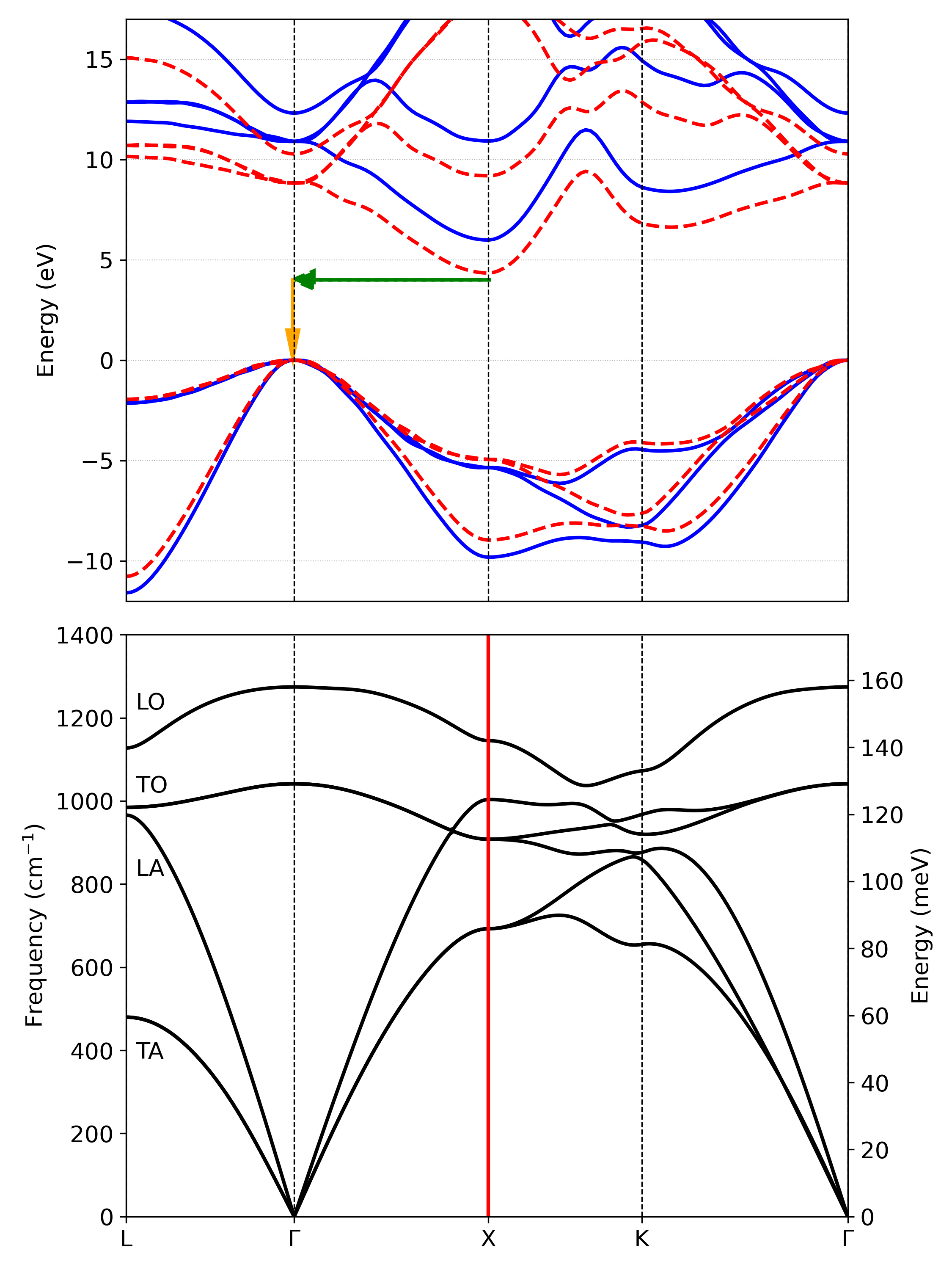}
		\caption{\label{fig:bands} In the top panel, we show the electronic band structure calculated at the KS (dashed red line) and G$_0$W$_0$ (solid blue line) approximation. In the same panel, we also show the lowest phonon-assisted transitions, with the green (phonon-emission) and orange (photon-emission) arrows. In the bottom panel, we show the phonon bandstructure, and indicate in red the phonons responsible for the lowest phonon-assisted transitions.} 
	\end{figure}

    The ground-state properties of cubic boron nitride are calculated within density functional theory (DFT), as implemented in the Quantum Espresso (QE) package\,\cite{Giannozzi_2017}. The experimental lattice parameter \CLA{of 3.615~\AA} reported in Ref.~\onlinecite{knittle1989experimental} is used throughout. Vibrational properties and electron--phonon matrix elements are obtained using density functional perturbation theory (DFPT)\,\cite{Giannozzi_2017}. 
	To study the electronic bands of $c$BN, we diagonalized the Kohn-Sham (KS) Hamiltonian and used MBPT\cite{reining} implemented in the Yambo code \,\cite{yambo-code} to correct the KS eigenvalues \RED{while keeping the wavefunctions fixed}. 
In this way we obtain the quasiparticle bandstructure within GW approximation, $\Sigma = G_0W_0$ \cite{reining}, with the Godby-Needs  plasmon--pole model.\cite{stankovski2011g}
    Starting from the quasiparticle bandstructure we proceeded to calculate the optical response function. To achieve this, we solved the Bethe-Salpeter Equation (BSE)\,\cite{strinati}, which includes both local--field effects and \CLA{screened} electron--hole \CLA{attraction}. The BSE is written on a basis of electron--hole transitions $t=(n_1 \kk_1 ) \rightarrow t'=(n_2 \kk_2 )$\cite{reining,PhysRevB.88.155113}  and  we adopt the Tamm-Dancoff approximation, including therefore only resonant transitions, $(v,\kk - \qq) \rightarrow  (c,\kk)$, 
    $v$ and $c$ denoting valence and conduction bands, respectively. The matrix elements of the BSE are:
		$\langle t' | H_{exc} (\qq) | t \rangle = E_t \delta_{t,t'} + \langle t' | \overline{V} -W | t \rangle
	$
    where $E_{t}$ is the quasiparticle energy associated with the above mentioned transition, $\overline{V}$ is the Coulomb potential and $W$ is the screened electron--hole interaction.\cite{strinati} 
    After diagonalization of BSE Hamiltonian we obtain the eigenstates $A^{\lambda}_{cv\boldsymbol{k}}(\qq)$ and energies $E_{\qq,\lambda}$ of the excitons. 
    The exciton energies and wave-functions at $\qq=0$ are used to build up the macroscopic dielectric function as:
	\begin{equation}
		\epsilon_M(\omega)= 1 - 4\pi\sum_\lambda \frac{ | T_\lambda|^2}{\omega - E_{\qq=0,\lambda} + i\eta},
		\label{eq:epsilon}
	\end{equation}
    where $T_\lambda =  \sum_{cv\kk} d_{cv\kk} A^\lambda_{cv\kk}(\qq=0)$ are the excitonic dipoles, $d_{cv\kk} = \langle v \kk | \hat r | c\kk \rangle$ are the KS dipole matrix elements and $i\eta$ is a small broadening term added to mimic the experimental resolution. 
	

	For the phonon-assisted absorption/emission, we follow the approach derived in Ref.~\onlinecite{PhysRevMaterials.7.024006} with two modifications. First, the phase of the exciton--phonon matrix elements has been fixed using the approach described in Refs.~\onlinecite{murali25arxiv,murali}. Second, the definition of phonon-assisted transition dipoles has been corrected to be gauge consistent. \CLA{This corresponds to including also the off-diagonal matrix elements of the exciton-phonon self-energy of Ref.~\onlinecite{PhysRevMaterials.7.024006}.}
    The first-order correction to the dielectric constant Eq.~\ref{eq:epsilon} induced by the coupling with phonons reads:
	\be
    \Delta \epsilon_M(\omega)=\sum_{ \mu\beta\qq} |\sum{_\lambda}  T_\lambda  \mathcal{D}^{\pm}_{\beta\lambda,\qq \mu}|^2 \frac{ F_{\mu \qq}^{\pm} }{ \omega - \left[E_{\qq \beta} \pm\,\omega_{\mu \qq }\right]  + i\eta }
	\label{eq:chi_lambda}
	\ee
    here, $\mathcal{D}^{\pm}_{\beta\lambda,\qq \mu}$ are the phonon-assisted \RED{satellite strengths}. 
    Eq.~\ref{eq:chi_lambda} describes the satellites generated by the emission or absorption of phonons by an exciton, and they appear as a peak at the energy of the finite-momentum excitons $E_{\qq,\beta}$ plus or minus one phonon energy $\omega_{\qq \mu}$. The `$\pm$' signs refer to phonon emission or absorption processes and \RED{$F_{\mu \qq}^{\pm}= 1/2 \pm 1/2 + N_{\mu, \qq}(T)$, with $N_{\mu\qq}(T)$ being the Bose-Einstein occupation function for phonons.} The appearance of satellites also induces a renormalization of the \RED{static} excitonic dielectric constant, a term that we do not consider here because the effect is small, and we investigate the two effects separately\cite{PhysRevMaterials.7.024006,marini2024optical}.
    The phonon-assisted \RED{satellite strengths} can be written explicitly as
	\bea
    \mathcal{D}^{\pm}_{\beta\lambda,\qq \mu}   &=& \frac{\mathcal{G}^{\mu \qq}_{\beta\lambda}}{(E_{\qq\beta} - E_{0\lambda} \pm \omega_{\mu \qq} + i\delta)}. \label{ph-ass-dip}
	\eea
    where $\mathcal{G}^{\mu \qq}_{\alpha\lambda}$ are the exciton--phonon coupling matrix elements at momentum $\qq$\cite{PhysRevMaterials.7.024006,marini2024optical}. 
    The small imaginary part $\delta$ in the \RED{satellite strengths} derives from the exciton--phonon self-energy\cite{PhysRevMaterials.7.024006}. We set its value to 1~meV, which is large enough to remove \CLA{unphysical} divergences and small enough not to affect the final optical spectrum. 

	
    To apply Eqs.~\eqref{eq:epsilon} and ~\eqref{eq:chi_lambda} to the light emission process, we use a steady-state approximation to obtain the luminescence via the \RED{``excitonic''}  van Roosbroeck-Shockley (RS) relation\cite{van1954photon,bbwill1972,Paleari2019}:
	\begin{widetext}
		\bea
		I^{(1)}_{PL}(\omega;T) \propto 
		\frac{1}{N_\qq}
		\sum_{\mu,\beta,\qq}
            \left|\sum_{\lambda}
			T^*_\lambda 	\mathcal{D}^{*,\mp}_{\beta\lambda,\qq \mu}  
		\right|^{2}
		N_{\beta \qq}(T_{\mathrm{exc}})
		F^{\pm}_{\mu \qq}(T)
		\delta\!\left(
		\omega - \left[E_{\qq \beta} \mp\,\omega_{\mu \qq }\right]
		\right).
		\label{lum_eq}
		\eea
	\end{widetext}
   \RED{In order to model this factor,} we suppose a steady-state condition where the exciton population can be approximately described by an equilibrium distribution that we assume to be a Boltzmann function, evaluated at an effective temperature $T_{\mathrm{exc}}$.  
	Finally, in a real luminescence calculation, we replace the delta function in Eq.~\eqref{lum_eq} by a Lorentzian, with a smearing set to 0.025~eV.

		\begin{figure}[ht]
		\centering
		\includegraphics[width=0.50\textwidth]{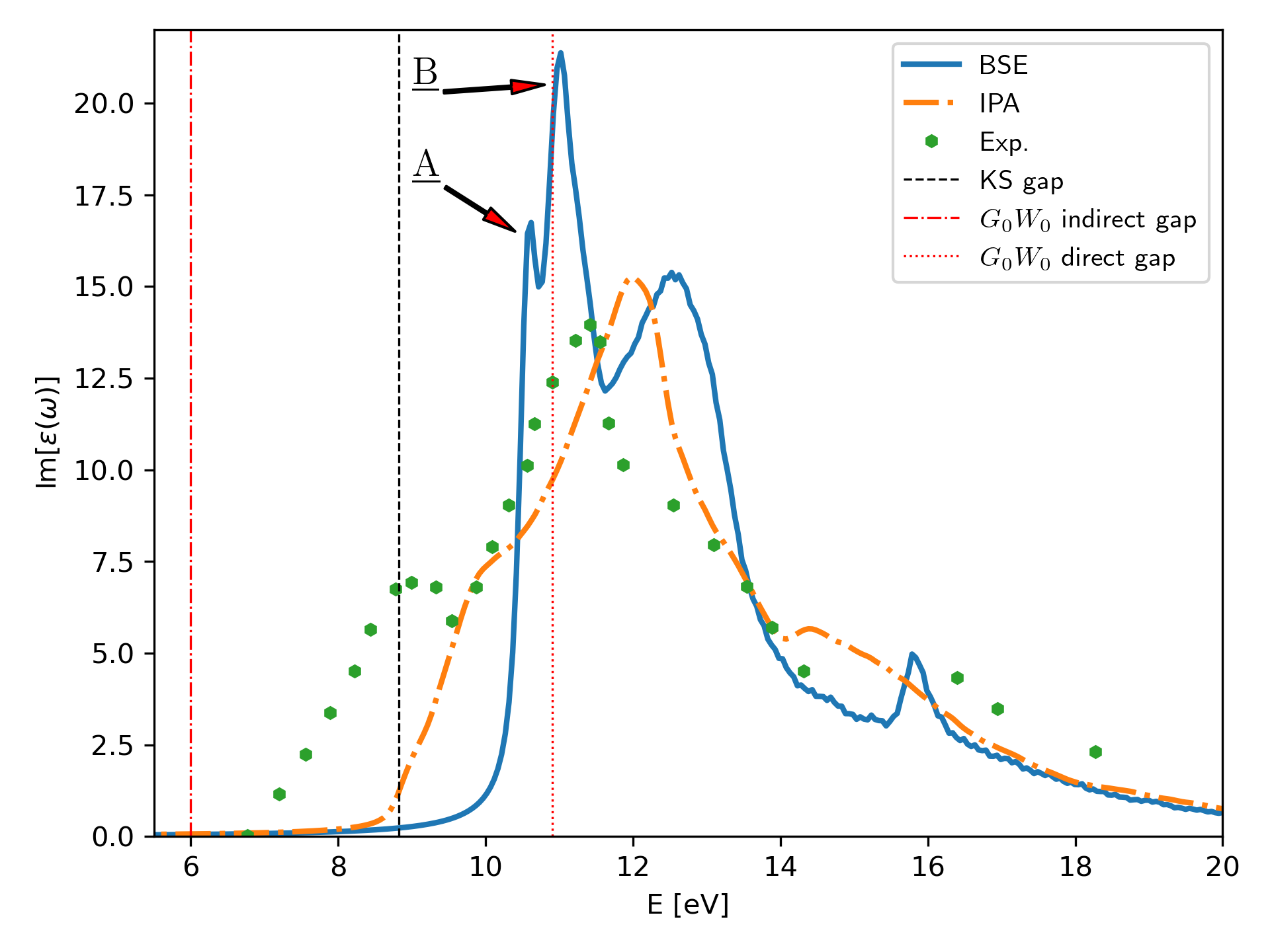}
		\caption{\label{bse_dbgrid} Optical absorption of cBN calculated at the BSE (solid blue line) and IPA (dot-dashed orange line) levels. \CLA{In the figure, we have indicated the first two peaks of the spectrum with A and B.} Results are compared with experimental measurements\cite{pouch1990synthesis}. In the same figure, we also report the direct KS (black dashed line) and the direct and indirect G$_0$W$_0$ gaps (red dash-dotted and dotted lines).}
	\end{figure}

	Now we \RED{briefly} present the computational details of the different calculations. 
Ground state properties and phonons were obtained using QE\cite{Giannozzi_2017} with DOJO pseudopotentials\cite{PseudoDojo}, PBE-sol functional\cite{perdew2008restoring}, 80~Ry cutoff, and $12\times 12 \times 12 $  k/q-grids.\\
    GW and BSE calculations were performed using the Yambo code\cite{yambo-code}. We used a cutoff of 6~Ha and 80 bands for the dielectric constant and the \RED{expansion} of the Green's function. In the BSE, we included 3 valence and 3 conduction bands. In agreement with previous studies\,\cite{satta2004many}, we found that the optical response of cBN requires a very dense grid to converge. 
    \RED{In order to accelerate it and obtain smoother spectra,} we used the double-grid method~\cite{alliati2022double} with the fine k-point grid of $61 \times 61 \times 61$ k-points \RED{in the calculation of the BSE optical absorption spectrum}. 
    Exciton--phonon coupling matrix elements were calculated using YamboPy\cite{yambopy} and the LetzElPhC code\cite{LetzElPhC} using the same band range and q-grid of the BSE. 
    The symmetry and selection rule analysis\cite{murali25arxiv} was also performed with Yambopy via the \texttt{spglib}\cite{spglib} and \texttt{MolSym}\cite{molsym} packages. \\
    In the calculation of phonon-assisted light absorption, exciton and phonon energies are interpolated on a large q-grid $61 \times 61 \times 61$ and used in the double-grid version of Eq.~\eqref{eq:chi_lambda} as described in Ref.~\onlinecite{PhysRevMaterials.7.024006}. In particular, phonon energies were interpolated using force constants\cite{Giannozzi_2017} while excitons were interpolated using a smooth Fourier interpolation.\cite{interpolation} For the calculation of the phonon-assisted dielectric constant we included the lowest 9 excitonic states and \RED{100} scattering states in the $\lambda$-sum of Eq.~\ref{eq:epsilon}, enough to converge the phonon-assisted absorption/emission in the luminescence spectrum. 
    We consider an excitonic and phononic temperature of 10~K and 300~K, corresponding to the occupation of the deep minimum of the exciton dispersion.

    Figure~\ref{fig:bands} (top panel) shows the bandstructure of cBN calculated at the KS and G$_0$W$_0$ level. We find the direct band gap at $\Gamma$ of 8.82 / 10.91~eV at DFT / G$_0$W$_0$ levels. The indirect gap between $\Gamma-X$ is 4.34~eV / 6.0~eV \RED{in the two cases}. 
    Surh et al.\cite{PhysRevB.43.9126} found 8.6 / 11.4~eV for the direct and 4.3 / 6.3~eV for the indirect gaps. 
    Satta et al. found 11.4~eV in the G$_0$W$_0$ direct case.\cite{satta2004many}. 
    Experiments report and 14.5~eV and $6.4 \pm 0.5$ for the two gaps\,\cite{PhysRev.127.159}. 
    In general we expect our GW correction to be smaller than the one obtained from Surh et al. due to the different plasmon--pole model\cite{stankovski2011g}, while regarding the comparison with Satta et al., two factors can explain the difference: first, we used the experimental lattice constant while they optimized it in DFT, and second, the parameters of our calculations are much more converged. 
    Regarding the experimental gap result, it must be considered with great uncertainty, given that it is a marginal result with respect to the main aim of that work. 
	
	In large band gap materials, G$_0$W$_0$ typically underestimates the gap correction; \RED{self-consistency on \RED{eigenvalues} improves experimental agreement\cite{shishkin2007self,artus}}. However, in this work, we decided to stick with the G$_0$W$_0+$BSE approximation in order to be consistent with other works in the literature and based on the assumption that any possible error in this approximation is of the same order in all BN polymorphs, so that we can compare our results with others on similar materials. 
\CLA{We also neglect all energy renormalizations due to electron-phonon interactions, namely quasiparticle band-gap reduction and exciton binding energy increase, which partially compensate.}
In summary, our results should be interpreted with the understanding that certain physical effects, \CLA{mostly pertaining to the absolute energy position of the optical spectra}, are not taken into account\CLA{, while they provide a good description of the optical response of cBN as far as coupling strengths and relative energy positions are concerned.}

	
    Figure~\ref{fig:bands} bottom panel shows the phonon bandstructure calculated with DFPT\cite{Giannozzi_2017}. With the same approach, we also calculated electron--phonon \RED{coupling matrix elements} that enter into the definition of the exciton--phonon matrix elements. \RED{For reference,} we found an LO-TO splitting at $\Gamma$ of 29~meV and the overall phonon bands are in good agreement with experimental measurements\,\cite{cai2007infrared,eremets1995optical} and previous calculations\cite{PhysRevB.63.115207}.
		

	\begin{figure}[ht]
		\centering
			\includegraphics[width=0.495\textwidth]{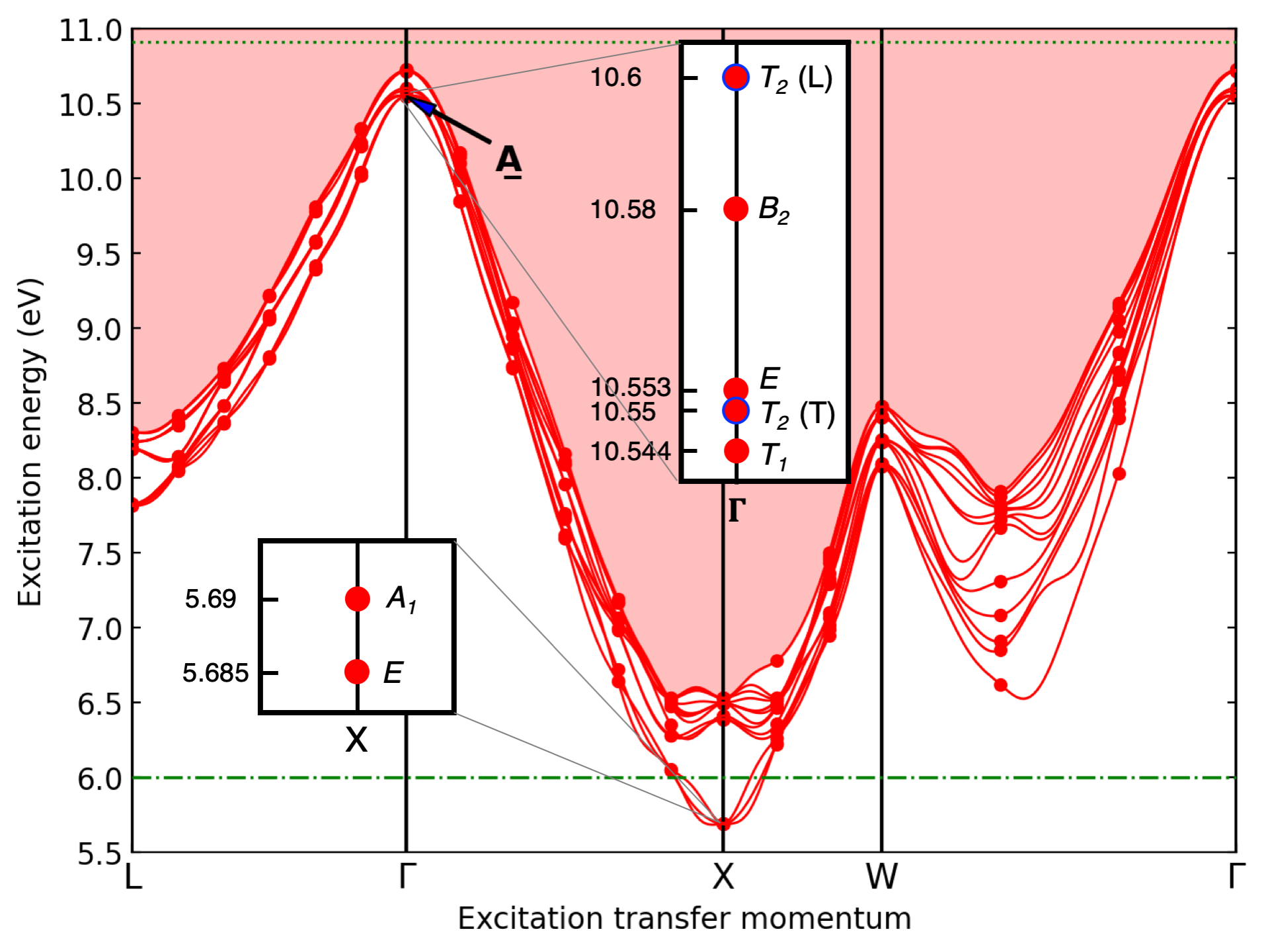}
            \caption{\label{exc_disp} Excitonic dispersion of cBN for the lowest 20 excitons interpolated using a smooth Fourier transform.\cite{interpolation} The position of the 3 times degenerate exciton state responsible for the first absorption peak $A$ at $\qq = \Gamma $ is marked, as well as the indirect and direct G$_0$W$_0$ gaps (green dash-dotted and dotted lines). \RED{The insets show a zoomed sketch of the two states at $\qq=\mathrm{X}$ (the global minimum) and the 9-state multiplet at $\qq = \Gamma $, including LT splitting of the bright exciton.}} 
	\end{figure}
		\begin{figure}[ht]
			\centering
			\includegraphics[width=0.5\textwidth]{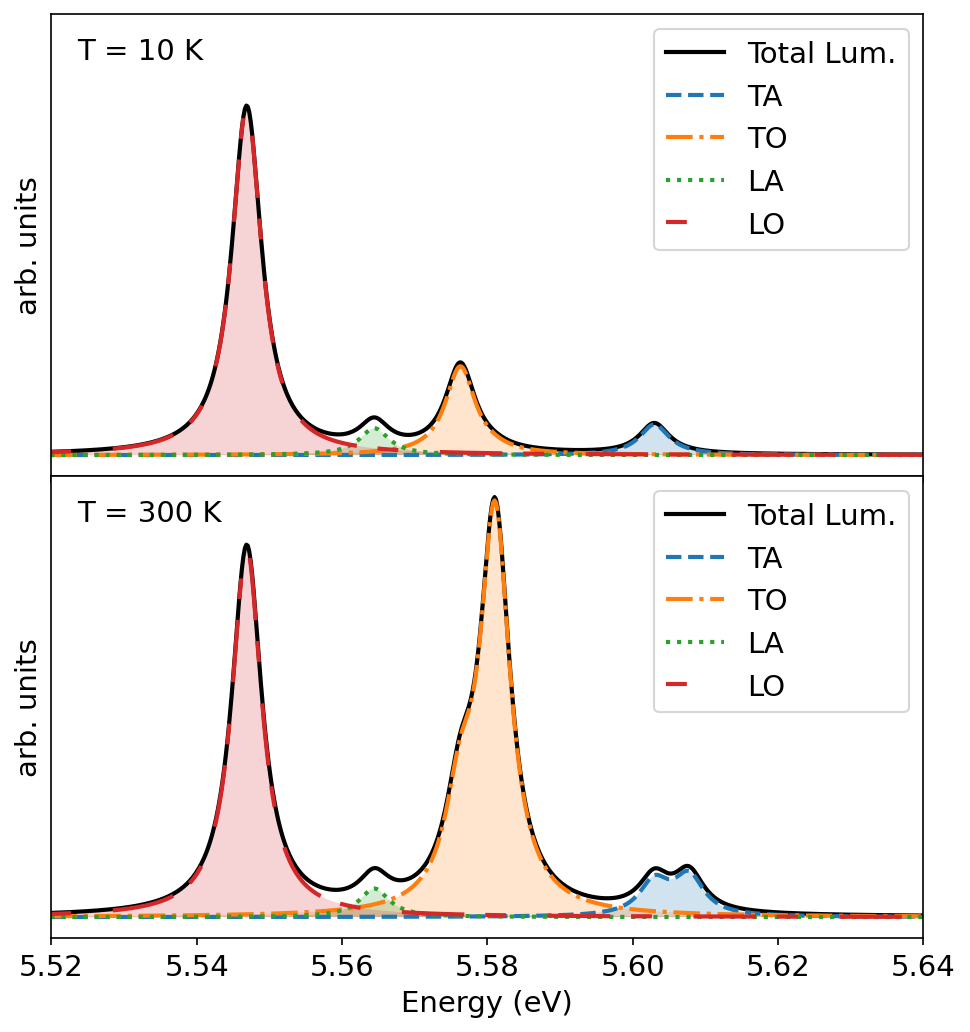}
			\caption{\label{fig:luminescence} Total phonon-assisted luminescence \RED{at 10 and 300~K} of cBN (continuous line) and the contribution of the different phonon \RED{branches} at $\qq=X$: LO (red loosely dashed line), LA (green dotted line), TO (orange dot-dashed line), TA (blue dashed line).} 
		\end{figure}
        Starting from the quasiparticle bandstructure calculated in the previous section, we solved the BSE. Then we calculated the optical absorption of cBN according to Eq.~\ref{eq:epsilon}. 
        In Fig.~\ref{bse_dbgrid} we report the BSE results compared with the independent particle approximation (IPA) starting from the KS bandstructure.
		
        We found that the onset of absorption begins at around 10.5~eV, at much higher energies than those reported in experiments. 
        In Fig.~\ref{bse_dbgrid} we have also indicated the first two absorption peaks with letters A and B, the first being a bound state with binding energy of approximately 0.3~eV and the second above the G$_0$W$_0$ \RED{direct} gap. 
        Compared to the IPA, the combination of GW+BSE shifts the spectrum, and several bound excitonic states form \RED{(many of them dark)}. 
        However, the situation is quite different from that of hBN, where all the spectral weight is concentrated in a few excitonic states\cite{paleari2018excitons}. 
        In the cBN case, the spectrum is the sum of many excitonic states with close energies. This is clearly visible when the excitonic dispersion is plotted. 
        To do this, we solved the BSE at finite momentum on a regular q-grid. 
        In Fig.~\ref{exc_disp}, we report the exciton dispersion along high-symmetry lines. The indirect nature of cBN is clearly visible: in particular, the $\Gamma$ point does not lie at the bottom of a valley, but instead is one of the highest energy points in the bands. 
        Here, we have a low-energy subset of 9 exciton states between 10.54 and 10.6 eV. The first absorption peak visible in Fig.~\ref{bse_dbgrid}, denoted as `A', lies in this subset: it is three times degenerate and transforms as the $T_2$ representation of the $T_d$ point group of cBN. 
        For reference, the longitudinal--transverse (LT) splitting of this state amounts to 0.05~eV (the longitudinal state jumps from 10.55 to 10.6 eV).
        There is also a dark $T_1$ triplet lying 15 meV below the A~peak, while the remainder of the 9-state subset is composed by a doubly degenerate state ($E$) and a nondegenerate one ($A_1$).
        The minimum of the excitonic dispersion is located at the $X$ point at $5.685$~eV, almost 5~eV below the $\Gamma$ point, consistent with the deep indirect nature of the electronic gap.
        Here, the minimum is formed by three states: first, a doubly degenerate state transforming as the $E$ representation of the little group $D_{2d}$ of the $X$ point, followed 5 meV above by a nondegenerate $A_1$ state.
		
\begin{figure}[ht]
			\centering
			\includegraphics[width=0.48\textwidth]{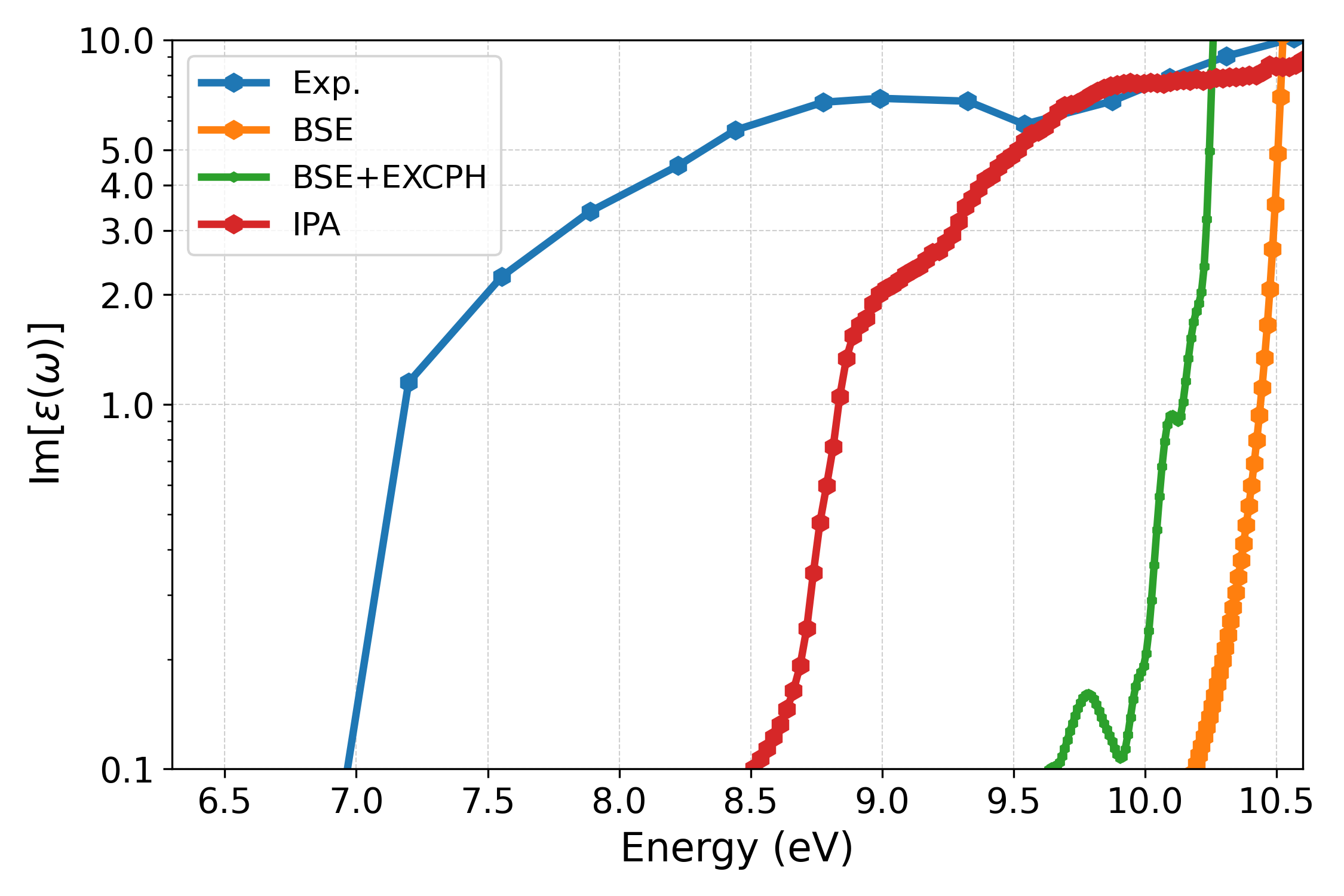}
                \caption{\label{fig:eps_ph} Low energy tail of the optical absorption of cBN at different levels of approximation: IPA, BSE, and BSE plus exciton--phonon coupling.  Results are compared with experimental measurements\cite{pouch1990synthesis}.}
		\end{figure}

        In Fig.~\ref{fig:luminescence} we report the phonon-assisted luminescence of cBN calculated using Eq.~\ref{lum_eq} at 10~K. At this temperature, the lowest-energy exciton ($E$) at $q=X$ is occupied, while the $A_1$ state above remains empty (the luminescence originates from the lowest excitons at the $X$ point). We found that the signal is centered around 5.55--5.6~eV, well below that of hBN (from 5.77–5.9 eV). In the same figure, we report the decomposition of the luminescence signal into the contribution of the different phonon modes. 
    To interpret the luminescence spectrum, we refer to the ${X}$ phonons shown in the lower panel of Fig.~\ref{fig:bands}, highlighted by a vertical red line in the same figure. These are the phonon modes that match the momentum of the lowest excitonic states in Fig.~\ref{exc_disp}. In order to distinguish the different phonon modes we  use the same nomenclature as for $\Gamma$ phonons, as it is clear from the continuous lines of the phonon dispersion which mode we are referring to. Combining all this information we found that phonon-assisted transitions are primarily mediated by transverse phonon branches with a small contribution of longitudinal branches at lower energies.
    For the $E$ exciton, it is important to emphasize that the selection rules for the phonon-mediated recombination process 
    allow for coupling with any phonon mode. Instead, in the case of the $A_1$ exciton, the coupling with the mode corresponding to the `LA' branch is forbidden.
    Compared to the more famous and better characterized luminescence spectrum of hBN\cite{Cannuccia2019,Paleari2019,PhysRevLett.131.206902}, the spectrum of cBN presents several differences: firstly, it is located at a lower energy; secondly, longitudinal and transverse phonons at $X$ are not so separate as in hBN to generate a doublet in luminescence; thirdly, unlike hBN, the phonon that contribute most are from \RED{the `LO' branches while `TO', `LA' and `TA' are lower in intensity}. All these characteristics will make it possible to distinguish it in future experiments on clean samples of cBN. 
    \RED{The bottom panel of Fig~\ref{fig:luminescence} displays the luminescence at 300 K. At this temperature, the $A_1$ excitons are thermally populated, and their recombinations are primarily mediated by the `TA' and `TO' phonons, each now generating a new luminescence satellite to the higher-energy side of the $E$ exciton ones.}
    \RED{Interestingly, the `TO' contribution from the A$_1$ exciton is more intense than the corresponding one from the $E$ state, pointing to a much larger exciton-phonon coupling for the former that is switched on above $T\sim 60$ K.} 
	Finally, in Fig.~\ref{fig:eps_ph} we report the low energy tail of the dielectric function including the contribution from the exciton--phonon scattering \CLA{obtained from Eq.~\ref{eq:chi_lambda} interpolated on a double-grid for exciton and phonon energies\cite{PhysRevMaterials.7.024006} at 10~K}. We found that coupling with phonons redshifts the onset from 10.3 to \RED{9.6}~eV. Looking at the excitonic dispersion, Fig.~\ref{exc_disp}, one would expect even a lower energy contribution of the epsilon. However, transitions toward low excitons close to $X$ have a very weak intensity due to the large denominators that appear in Eq.~\ref{ph-ass-dip}. For this reason, the first visible contribution to the epsilon is only 1 eV below the direct transitions. 

		In conclusion, we have investigated the optical response of cubic boron nitride (cBN) using first-principles many-body perturbation theory, specifically accounting for the critical role of exciton--phonon coupling. 
        Our findings provide a better understanding of the long-standing discrepancy between theoretical predictions of a $\sim 11$~eV optical gap and experimental reports of absorption and emission features in the 6-7 eV range.  
        We have shown that while the direct optical gap remains high, the inclusion of exciton--phonon interactions effectively lowers the observable absorption onset. 
	However, intrinsic exciton--phonon coupling in the pristine system is not enough to reconcile theory with experimental observations. In fact, even if excitons at very low energy are present in cBN, their contribution to the dielectric constant is suppressed by the large denominators in the phonon-assisted dipoles. We expect that the low-energy contribution in the experimental measurement of the dielectric constant of cBN is due to the presence of defects, through direct transitions to defect levels or to the activation of transitions to finite momentum excitons due to the breaking of translation symmetry\cite{tararan2018optical}.\\
        Regarding the luminescence, our results show that cBN and hBN exhibit distinct spectral profiles. We found that the intrinsic phonon-assisted luminescence of pure cBN is centered around 5.57--5.60~eV, which is lower than that typically observed for hBN, from 5.77--5.9~eV. This energy shift suggests that experimental emission signals reported around 6 eV may be attributed to hBN inclusions within cBN samples rather than the intrinsic cBN or defect--related states. 
	This underscores the sensitivity of luminescence spectroscopy as a tool for phase characterization in boron nitride polymorphs. 
		
		
		\begin{acknowledgments}
                C.A. and A.P. acknowledge B. Demoulin and  A. Saul for the management of the computer cluster \emph{Claudia}. F.P. acknowledges M. Nalabothula for insightful discussions on exciton symmetries and luminescence. C.A. acknowledges funding from European Research Council MSCA-ITN TIMES under grant agreement 101118915. E.C. acknowledges ANR project  COLIBRI No. ANR-22-CE30-0027. C.A., A.M. and A.P. acknowledge support from the Amidex foundation through the project INDIGENA. F.P. Acknowledges support from ICSC-CNR in High Performance Computing, Big Data and Quantum Computing - funded by the European Union(EU) through the Italian Ministry of University and Research under PNRR M4C2I1.4 (Grant No. CN00000013). This work has also been supported by the EU through the MaX ”MAterials design at the eXascale” Centre of Excellence (Grant agreement No. 101093374, co-funded by the EuroHPC JU).
		\end{acknowledgments}
		\bibliographystyle{apsrev4-1}
		\bibliography{wBN,cBN}
		
	\end{document}